# Modeling and predicting measured response time of cloud-based web services using long-memory time series

Hossein Nourikhah · Mohammad Kazem Akbari · Mohammad Kalantari

**Abstract** Predicting cloud performance from user's perspective is a complex task, because of several factors involved in providing the service to the consumer. In this work, the response time of 10 real-world services is analyzed. We have observed long memory in terms of the measured response time of the CPU-intensive services and statistically verified this observation using estimators of the Hurst exponent. Then, naïve, mean, autoregressive integrated moving average (ARIMA) and autoregressive fractionally integrated moving average (ARFIMA) methods are used to forecast the future values of quality of service (QoS) at runtime. Results of the cross-validation over the 10 datasets show that the long-memory ARFIMA model provides the mean of 37.5 % and the maximum of 57.8 % reduction in the forecast error when compared to the short-memory ARIMA model according to the standard error measure of mean absolute percentage error. Our work implies that consideration of the long-range dependence in QoS data can help to improve the selection of services according to their possible future QoS values.

H. Nourikhah · M. K. Akbari (✉)
Department of Computer Engineering and Information Technology, Amirkabir University
of Technology, 424 Hafez Ave, 15875-4413 Tehran, Iran
e-mail: akbarif@aut.ac.ir

H. Nourikhah
e-mail: nourikhah@aut.ac.ir

M. Kalantari
Department of Electrical and Computer Engineering, Shahid Rajaee Teacher Training University,
Tehran, Iran
e-mail: mkalantari@srttu.edu

## 1 Introduction

In a cloud computing environment with service-oriented architecture, it is possible to use services that are implemented by different vendors to create distributed enterprise applications that scale-up very well. This architecture is based on a distributed element called service that uses standard and well-defined messages in extensible markup language (XML) format. Services communicate in a loosely coupled manner and collaboratively perform tasks. In fact, this relaxation from knowing service implementations helps service providers to create composite services with arbitrary implementations. It is even possible to use several instances of a service to improve the availability.

E-commerce websites widely use web services on the Internet. When a customer orders goods to an E-commerce website, there is a requirement to check his/her credit card data. Confirming this data is usually accomplished with the help of a third party that is a service card issuer. Upon confirming purchases, the providing company contacts the transportation company to handle the shipment of the goods and deliver them to the customers. This process can be defined and controlled easily in a service-oriented architecture, and it is possible to control the performance of the process and identify low performance subsets of the process or actors that cause difficulty in accomplishing the whole task.

Services can be used either with static or runtime binding. In the first mentioned approach, usable services are pre-defined and composition is statically created and used. Although this approach can perform the computations in a distributed manner, it has several shortcomings. For example, if one service fails at runtime, there will be no way to fix the composition other than waiting for the failed service to be repaired.

A better approach is runtime binding, in which it is possible to find several candidates to perform a single task and then use the best candidate at runtime. This approach comes from standardization efforts within the IT industry. With the standardization of operations and data formats, several services can be created by different vendors that provide similar functionality. In this way, it is possible to choose better services from candidates that provide desired functionalities to achieve the best possible quality. To achieve such a goal, we need to log the quality of previously invoked services and use it as a judgment basis for selecting better services from the available candidates.

Quality of service consists of several parameters that are beneficial for the customers. The exact parameters are different according to the context, but some of the general parameters are availability and performance measures, such as the response time and bandwidth. Because the value of quality of service (QoS) comes from the past, methods are needed to calculate the effective values at runtime. QoS values change over time because of the dynamism in the environment; thus, a service that is selected based on current QoS value can have shortcomings at runtime.

In this work, the response time of 10 real-world services is analyzed. An important parameter in the time series is the memory of it. The autocorrelation of measured data shows that the response time in CPU-intensive web services is a long-range dependence time series.

Stationary process $X(t)$ is said to have long-memory or long-range dependence [2] if its sample autocorrelations, defined as $\hat{\rho}(k) = \hat{\gamma}(k)/\hat{\gamma}(0)$, decay to zero at a rate approximate to $k^{-\alpha}$ for some $0 < \alpha < 1$. We use $H = 1 - \alpha/2$, thus if the estimated value conforms $1/2 < H < 1$, we have a long-memory process.

Careful statistical investigation of measured QoS data of real-world web services is the main contribution of our work, with the goal being to better understand QoS data, which can lead to better identification of patterns and eventually to achieve better predictions.

The remainder of the article is organized as follows. In Sect. 2, we provide background and related works. In Sect. 3, we discuss time series models, including the short-memory model autoregressive integrated moving average (ARIMA) and the long-memory model autoregressive fractionally integrated moving average (ARFIMA). In Sect. 4, we discuss the dataset used for our experiments, and then we provide the results of our extensive experiments. Finally, we provide conclusion and some discussion for future works in Sect. 5.

## 2 Related studies

Time series analysis has a wide range of applications in science, and specifically self-similar (fractal) time series are widely used in different scientific areas including hydrology [9], geology [15], econometrics [18], physics [25], computers, communications, and many science fields that study nature.

Self-similarity is a feature that is found in several natural phenomena. When looking at a fractal such as the Koch snowflake or Mandelbrot set, we see approximately the same thing that we see by increasing the scale and zooming into the fractal. This feature exists in many natural time series. For example, in 1965, Hurst [9] found this feature in rain and drought conditions of the Nile river flow.

Although studies on self-similarity and fractals dates back to the seventeenth century, among the studies of recursive features, Mandelbrot [16] defined self-similarity and fractals in an exact manner using rigorous statistics and showed its application to various aspects of science. This mathematician defined fractals as having the feature of self-similarity and scale-invariance. The shape of scale-invariant objects does not change much on different scales, and it is not possible to infer the scale by looking only at the shape. In his paper, Mandelbrot introduced these properties using the length of the coast of Britain. In fact, self-similarity is widely found in nature [17], and several different phenomena can be described using self-similarity. Many research studies have been accomplished in several areas while considering self-similarity, and we examine only the related studies in the area of communications.

Mathematically speaking, Stochastic process $X_t$ is considered self-similar with self-similarity parameter $H$ [2] if for any positive stretching factor $c$, the rescaled process with time scale $ct$, $c^{-H} X_{ct}$ is equal in distribution to the original process $X_t$.

Self-similar time series with long-range dependence are very useful in computer and telecommunications, and are used for more realistic modeling of the Internet [22]. Leland et al. [13,14], by investigating the empirical results from carefully measuring Ethernet traffic with a special custom-designed measurement device in a 4-year period showed that the nature of Ethernet traffic is self-similar and long-range dependent and that it is bursty at different time scales. This feature is not present in a Poisson traffic model that was shown to be suitable for modeling telephone networks and was previously used in computer networks.

With these findings, and by considering the large effect of self-similar traffic on the fast filling of routers' buffers, it became clear that Poisson traffic is not suitable for modeling network traffic in wide area networks [23]. Because of long-range dependence (LRD), it is inevitable to revise network analytical models for network traffic, find better traffic generation and simulation methods, re-evaluate network models and measure their performance with fractal traffic.

LRD causes problems for the results that are obtained using queue networks in analyzing routers and switches because most of the pre-LRD results were obtained using the assumption of a Poisson traffic model, and the calculated buffer and other parameters would not suffice in real-world LRD traffic. Thus, new extensions of network calculus are used to determine the effects of self-similar traffic. In this research, we do not intend to improve the design and/or restructure the routers and switches, but we want to have a good selection among the available web services. Thus, we want to predetermine the status of Internet services according to the history of their response time.

Self-similar traffic has a serious impact on needed temporary buffer and other attributes of routers and switches. Thus, the necessity of considering self-similarity and providing models for investigating and forecasting these attributes is clear. Two main features of self-similarity are the existence of large changes in a short period of time (large variance) and the persistence of a feature over a long period of time and then the disappearance of it. These two features were called the Noah and Joseph effect, respectively, by Mandelbrot and Wallis [19] because of their similarity to historical events. These effects are also available in Internet traffic time series. According to statistical research on the causes of self-similarity, the root of this feature in Internet traffic is that aggregating several ON/OFF sources with the Noah effect will create aggregate traffic with the Joseph effect [32].

One of the most important features of self-similar traffic is that it preserves its bursty feature at different scales. Statistically, this attribute means that self-similar time series have the same statistical features at different time scales which range from seconds to minutes or even hours. In addition, aggregating self-similar traffic does not result in smoothing.

Previous studies in the area of Internet performance modeling can be categorized into two separate categories (as shown in Fig. 1): traffic modeling and QoS modeling using empirically measured data. Although these two categories have many things in common, they have different goals. In the first category, which is traffic modeling, the goal is usually to attain better modeling of traffic to improve the infrastructure and create better routers, switches and other network equipment.

In the second category, the modeling of QoS, the goal is to predict the future states of services to be able to select better services. Research in this category is usually applicable where the goal is not to change the infrastructure, or it is not easily possible to change it, but it is possible to choose between various services. In this case, it is very important to know the dynamism of the service quality in the available services to be able to choose service that will have better quality at runtime.

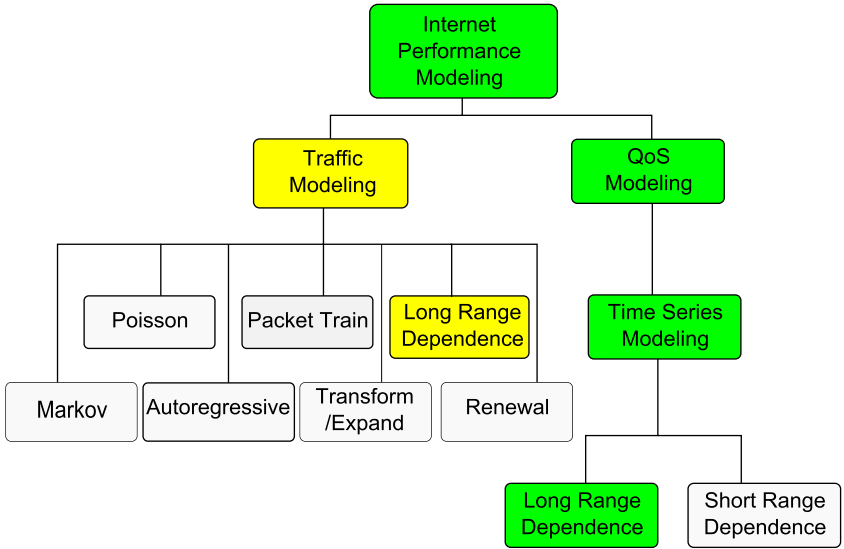

**Fig. 1** Taxonomy of research on Internet performance modeling

Using time series models to forecast future values of QoS has been considered by different researchers in recent years. However, it is worth noting that none of these studies have examined the memory of time series, and they have used common short-memory models to forecast future values.

Cavallo et al. [4] gathered empirical QoS data by monitoring 10 web services over a period of 4 months. They have collected more than 3,000 samples for each service, including the time of request, size of request, response time, and success or failure of the service invocation. Then, using this empirical data, they used several different forecasting methods, including the last value, mean and ARIMA to calculate possible future values of the response time of the web services. This research showed that ARIMA provides good forecasts, considering the error variance. In the mentioned research, kernel smoothing is used to reduce outliers, but our work shows that large values in QoS data do not necessarily indicate errors in measurements, and they can come from self-similarity as the underlying feature of the dataset. Thus, kernel smoothing can create worse forecasts.

Solomon et al. [28] have tackled the problem of predicting future QoS values by considering key performance indicators (KPIs) and predicting them using the ARIMA method. The difference of their work from the previous work is the aggregation of QoS data in time windows and then working on aggregated data as a key performance indicator. In this work, ARIMA and linear regression are compared according to the different error metrics mean absolute scaled error (MASE) and weighted absolute percentage error (WAPE), which show the mean and worst error, correspondingly. In each time window, an ARIMA model is created for each QoS parameter, and the best model is used. Their results show that ARIMA provides better forecasts for aggregated data.

Considering the volatility in the above-mentioned dataset, Amin et al. [1] used methods that are suitable for non-stationary methods and suggested using generalized autoregressive conditional heteroskedasticity (GARCH) and self exciting threshold autoregressive fractionally moving average (SETARMA).

Leitner et al. [12] provided a framework that attempts to minimize QoS violation penalties by exploring QoS history data of several services in a composition and providing a good forecast. The mentioned research is one of the newest applications in the area of web service QoS assurance and uses different prediction methods at different levels of service level objectives (SLO) and especially the SRD model ARIMA to predict the quality of composite web services at the level of Aggregation.

Thus, previous studies have ignored the important feature of "self-similarity" in QoS data, which provides better knowledge about the nature of the data that can eventually lead to better forecasts. In our work, with careful statistical investigation, we show that measured response time of CPU-intensive services is self-similar and has long-range dependence.

To fit and forecast long-memory time series, various models are proposed, among which are ARFIMA, Fractionally Integrated generalized autoregressive conditional heteroskedasticity (FIGARCH), and Gegenbauer autoregressive moving average (GARMA). We use ARFIMA to forecast the measured response time values, and we compare this approach with other methods such as the mean, last value, and ARIMA to show that the LRD model ARFIMA can give better forecasts.

Granger and Joyeux [8] pointed that "models using fractional differencing will not necessarily provide clearly superior short run forecasts, but they may give better longer-run forecasts where modeling the low-frequencies properly is vital". Results of our experiments verify this hypothesis.

## 3 Modeling using time series analysis

Predicting future values of QoS with a small error margin can help in choosing better services. One of the useful methods for achieving forecasts is time series analysis. This method is widely used in econometrics, geological modeling, signal processing, weather forecasting, and several other scientific contexts.

Various models are usable in time series analysis, but not all of them are necessarily good for every type of time series. Thus, it is important to use time series models that correctly specify the underlying data-generating process. To show that a specific model fits well on a specific dataset and provides good forecasts, one should estimate parameters that fit the model on the dataset and then choose a model that provides the best forecasts and the least error according to standard error criteria.

### 3.1 Time series

First, we introduce the notations that are used in the remainder of the paper, and then, several models, including naïve, linear regression, mean, Autoregressive (AR), Moving Average (MA), ARMA, ARIMA and suitable models for long-memory time series and the case of ARFIMA, are discussed.

Considering a QoS variable such as $X$, the observed value at time $t_i$ $(0 \leq i \leq k)$ is denoted as $\hat{X}(t_i)$. So, the time series of observed values is denoted as:

$$\hat{X}(t_0), \hat{X}(t_1), \cdots, \hat{X}(t_k) \tag{1}$$

It is obvious that the predicted values of the time series are not exactly equal to the real observed values. Thus, we show the values of the time series $X$ at time $t_i$ by $X_i$. To forecast future values, we first need to fit a time series model on the measured values of the past.

### 3.1.1 Naïve

The easiest possible method is to use the last observed value (naïve). In many existing systems, the last measured value of the QoS is used for the selection of services in the future (next $h$ time steps):

$$X_{k+n} = \hat{X}(t_i) \tag{2}$$

This method is very simple, but it ignores values other than the last value, which can lead to inaccurate results and wrong service selections.

### 3.1.2 Mean

Using the mean of previously measured values is another simple method that is often used in forecasting:

$$X_{k+n} = \sum_{i=0}^{k} \frac{\hat{X}(t_i)}{k+1} \tag{3}$$

This method is less error-prone compared with the naïve method because it accounts for all of the previously measured values in the future estimation. However, it has some drawbacks, such as losing the variance. In this method, the variance of the previous values is not presented, and the fluctuations are not modeled correctly [5]. It is useful to mention that every other method that is used is typically compared with the mean because if a method performs worse than the mean in forecast accuracy, it most often uses more complex calculations, and thus, it is not worth using [10].

### 3.1.3 ARIMA models

In autoregressive models, the variable that is forecasted is calculated using a linear combination of the values of the same variable. If $\{Z_t\}$ is white noise, then $\{X_t\}$ is called AR($p$) or Autoregressive of degree $p$ if [3]

$$X_t = \varphi_1 X_{t-1} + \cdots + \varphi_p X_{t-p} + Z_t \tag{4}$$

in which the $\varphi_i$'s are constants. In this formula, $X_t$ is not regressed on independent variables and instead is regressed on older values of the same variable $X_t$. If $Z_t$ is a random variable that has a mean of 0 and a variance of $\sigma^2 z$, then $X_t$ will be MA($q$) if

$$X_t = Z_t + \beta Z_{t-1} + \cdots + \beta_q Z_{t-q} \tag{5}$$

where the $\theta_i$'s are constant. This equation can be written in simpler form using a backward shift operator $B^j X_t = X_{t-j}$ as $X_t = \Theta(B)Z_t$, in which $\Theta(B) = 1 + \theta_1 B + \cdots + \theta_q B^q$.

In other words, MA($q$) describes the current value according to $q$ previous pulses, and AR($p$) describes the current value according to $p$ previous events. Next, it is possible to define ARMA($p, q$) as a combination of these two:

$$X_t = \varphi_1 X_{t-1} + \cdots + \varphi_p X_{t-p} + Z_t + \theta_1 Z_{t-1} + \cdots + \theta_q Z_{t-q} \tag{6}$$

To use the ARMA model, first we need to ensure stationarity. A stationary time series is a time series for which the statistical properties do not depend on the time frame that the series is observed in. The unit root test is one of the objective tests for stationarity; it tests for the differencing that is needed to achieve a stationary time series. Several unit root tests are available that by different assumptions lead to different results. The most useful of such tests are the augmented Dickey Fueller (ADF) and Kwiatkowski–Phillips–Schmidt–Shin (KPSS) test.

Here, we use the ADF test on which this model is estimated by

$$y'_t = \varphi y_{t-1} + \beta_1 y'_{t-1} + \beta_2 y'_{t-2} + \ldots + \beta_k y'_{t-k}. \tag{7}$$

in which $y'_t$ is the first difference time series, $y'_t = y_t - y_{t-1}$. $k$ is the number of lags that are included in the regression. If the original $y_t$ needs differencing, then the $\hat{\varphi}$ coefficient would be near to zero, and if $y_t$ is stationary, then it would yield $\hat{\varphi} < 0$.

The null hypothesis in the ADF test is that the data are non-stationary. Thus, the large values of $p$ show non-stationarity, and the small $p$ values show stationarity in the time series. Usually, a 5 % threshold is used, and when $p$ is larger than 0.05, differencing is considered to be necessary.

If a times series is non-stationary, then it is possible to use the ARIMA($p, d, q$) model by transforming it into a stationary time series using differencing. This arrangement means that we transform the original time series by differencing $d$ times, into a secondary stationary time series. Thus, $W_t = \nabla^d X_t = (1 - B)^d X_t$ and $W_t$ can be written as

$$\begin{aligned} W_t &= \varphi_1 W_{t-1} + \cdots + \varphi_p W_{t-p} + Z_t + \theta_1 Z_{t-1} + \cdots + \theta_q Z_{t-q}, \\ Z_t &= WN(0, \sigma^2). \end{aligned} \tag{8}$$

Then, we perform model fitting and estimate the parameters according to certain criteria. In the ARIMA($p, d, q$) model, we first calculate $d$ by applying the difference operator several times to achieve a stationary process. Then, to estimate $p$ and $q$, according to the Akaike information criterion with correction for finite sample sizes

(AICc) criterion, we test different values up to a certain limit. AICc is defined as [3]

$$\text{AICc}(\varphi, \theta) = \frac{-2 \ln L(\varphi, \theta, S(\varphi, \theta)/n) + 2(p + q + 1)n}{n - p - q - 2} \quad (9)$$

where $\varphi$ and $\theta$ are calculated using the maximum likelihood method. One of the features of AICc is that it considers a penalty for larger values of $p$ and $q$, which means that large values of these parameters are avoided. After finding $p$, $d$ and $q$, the model parameters should be estimated again using AICc.

*3.1.4 ARFIMA model (time series with long memory)*

Models that we previously discussed have short memory. Working with long-memory time series, we should use the other forecasting models that are discussed here. Three categories of such models are ARFIMA, GARCH, and GARMA. Here, we use ARFIMA for modeling response time of Internet web services.

ARFIMA models were introduced by Granger and Joyeux [8]. In this model, fractional differencing is used to achieve an ARMA process. The ARFIMA process is defined as [21]:

$$\varphi(B) X_t = \theta(B) (1 - B)^{-d} \varepsilon_t \quad (10)$$

where $\varphi(B) = 1 + \varphi_1 B + \cdots + \varphi B^p$ is the AR coefficient, $\theta(B) = 1 + \theta_1 B + \cdots + \theta_q B^q$ is the MA coefficient, and these two have no common roots. $(1 - B)^{-d}$ is the fractional differencing operator, which is defined by

$$(1 - B)^{-d} = \sum_{j=0}^{\infty} \eta_j B^j = \eta(B) \quad (11)$$

in which the $\eta_j$ function is defined as

$$\eta_j = \frac{\Gamma(j + d)}{\Gamma(j + 1) \Gamma(d)} \quad (12)$$

where $d < 1/2$, $d \neq 0, -1, -2, \cdots$, and $\{\varepsilon_t\}$ is white noise with a finite variance. In ARFIMA process $\{X_t\}$, for $0 < d < 1/2$, the process has a long memory, and for $d = 0$, it has a short memory. This model has been used for Internet traffic forecasting [7,27] but not for QoS forecasting, for which we show here that LRD modeling is applicable.

ARFIMA processes which have a seasonal part are called Seasonal autoregressive fractionally integrated moving average (SARFIMA) processes. A general category of Gaussian processes that have a long memory and seasonal part can be defined according to their spectral density, as follows [21]:

$$f(\lambda) = g(\lambda) |\lambda|^{-\alpha} \prod_{i=1}^{r} \prod_{j=2}^{m_i} |\lambda - \lambda_{ij}|^{-\alpha_i} \quad (13)$$

where $\lambda \in (-\pi, \pi]$, $0 \leq \alpha, \alpha_i < 1, i = 1, \ldots, r$ and $g(\lambda)$ is a symmetric function that is strictly positive, continuous and bounded. There are other assumptions with respect to the symmetry of $f$ that we omit here.

SARFIMA models have a useful attribute that can be used to identify them. A plot of these processes has local minima at the frequencies $\lambda = 2\pi j/s$, $j = 0, 1, \ldots$, which is useful for identifying processes that have a seasonal part according to their ACF. A similar feature exists in PACF for the LRD time series that have a seasonal part.

However, as we show in the remainder of this paper, sometimes the intense fluctuations of the time series are such that we can ignore the seasonality compared with those large changes and consider the whole series to be stationary. In these cases, stationary tests are useful, and if they imply stationarity, we can ignore that little seasonal part and use a non-seasonal model of ARFIMA for the modeling.

### 3.2 Identifying long-range dependence

The LRD feature can be described as the persistence of autocorrelation for a long time, and we use the Hurst exponent as a measure to show the existence of self-similarity and long-range dependence. This parameter is usually shown by $H$ and was introduced by Hurst [9].

Not all self-similar time series are LRD. However, by looking at a Hurst exponent estimate, we can make a judgment about the existence of short or long memory in the time series: self-similar time series for which $1/2 < \beta < 1$ are LRD, and those in which $0 < \beta \leq 1/2$ are SRD. Several methods are proposed for estimating the Hurst exponent [26,33], and three of these methods were used here as follows.

#### 3.2.1 Rescaled range

The most famous method for estimating the Hurst exponent is the main method of Harold Edwin Hurst, which appeared in his original work in the area of Hydrology and was accomplished by observing the Nile River [9]. Hurst worked on flood and drought periods of the Nile for a long period of time, and by considering the flow of the river as a time series, he determined the level of water in an ideal reservoir. Based on his studies on the Nile River, Hurst defined rescaled range ($R/S$) analysis, which is more precisely defined in the works of Mandelbrot and Wallis [20].

We consider the time series $X_n$, $n = 1, \ldots, N$ and calculate the sample standard deviation $S_N$ and the mean-adjusted rescaled range

$$y_m = \sum_{n=1}^{m} (x_n - \bar{x}_N) = \left(\sum_{n=1}^{m} x_n\right) - m\bar{x}_N \tag{14}$$

The range is defined as

$$R_N = \max_{1 \leq m \leq N} (y_m) - \min_{1 \leq m \leq N} (y_m) \tag{15}$$

and the rescaled range is defined as $R_n/S_n$ in the following way

$$E\left[\frac{R_n}{S_n}\right] = Cn^H \quad (16)$$

For estimating the Hurst exponent, the time series is divided into $n$ parts in which $n = N, N/2, N/4, \ldots$ and $R_n/S_n$ is calculated for all of the values of $n$. Then, the mean is calculated over these values. After that, $H$ is calculated using a power law. This process can be accomplished graphically by drawing $E[R_n/S_n]$ in a log–log plot and fitting a straight line on the graph, and then calculating the slope of the line.

### 3.2.2 Periodogram

In this method, the Hurst exponent is estimated according to the spectral density of the time series. When the series has long memory, the variance is infinite, and the periodogram of the time series will conform to a power law. Thus, if the spectral density of the time series in the frequency domain is shown in a log–log plot, it is expected to be linear. In this way, the slope of the line will be $1 - 2H$, and $H$ can be estimated using this theoretical relation. In practice, working on the lower 10 % of the frequencies is sufficient, because only frequencies that are closer to 0 conform to the power law. For estimating the Hurst exponent by this method [30], we first calculate

$$I(\lambda) = \frac{1}{2\pi N}\left|\sum_{j=1}^{N} X_j e^{ij\lambda}\right|^2 \quad (17)$$

in which $\lambda$ is the frequency, $N$ is the number of elements in the time series, $X_j$ is the dataset, and $I(\lambda)$ is the spectral density. Thus, the LRD time series agrees with the power law in the frequency domain and will have a periodogram that is proportional to $|\lambda|^{1-2H}$.

### 3.2.3 Aggregated variance

In this method, the Hurst exponent is estimated from the variance of an aggregated time series. First, the original time series $X_t$ is divided into blocks that have the initial size of $m = 1$, and after that, in each subset, we calculate the mean as follows [26,33]:

$$X^{(m)} = \frac{1}{m}\sum_{i=(k-1)m+1}^{km} X(i), \ k = 1, 2, \ldots \quad (18)$$

Then, we calculate the variance of $X^{(m)}(k)$. In fact, the time series is divided into blocks that have the length of $m$, and then, the variance is calculated for each block. The slope of the line $(\beta)$ is calculated by fitting a straight line on a log–log plot of the sample variance against the block size by the least squares method. Then, $H$ is estimated according to the theoretical relation $\beta = 2H - 2$.

## 4 Data and results

### 4.1 Exploring data

The data that are used in this research are the measured QoS of 10 real Internet web services over a period of four months. In this study, we focus on the response time of the services. The used dataset was gathered by Cavallo et al. [4]. In Table 1, which is extracted from that research, all 10 of the probed web services are defined, and a description of each service is provided. In this study, we have analyzed these data with careful statistical measures. All of the statistical works were performed using the R environment for statistical computing [24]. We used the *forecast* [11] package for performing the forecasts, *fracdiff* [6] for LRD modeling, and *plyr* [31] for data aggregation.

Figure 2 shows the response time from the measured QoS data from the 10 web services. Note that, the measurement period is 1 h for all of the series except for services 4 and 7, which have the rates of 2 h and 30 min, respectively. From the graphs, there are specific features that are visible in the data, as follows; these visible features are then carefully investigated using autocorrelation and estimation of Hurst exponent.

1. There are several bursts in the data that are not confined to a specific period of time. This is visible in all of the series.
2. By examining the data at longer scales with shorter periods of time, it appears that the shape of the graph is similar to a graph with larger scales and longer periods of time. This visible feature—as defined before—is called self-similarity, and its existence is investigated in the following section by estimating Hurst exponent for each series. Except the series 3, 5, and 10, other series have this feature.
3. There is a strong correlation between the consecutive values. This is investigated later by examining the autocorrelations of time series, and it is shown that again only series 3, 5, and 10 lack this feature.

**Table 1** Description of the monitored web services

|    | Name                | Description                                     |
|----|---------------------|-------------------------------------------------|
| 1  | Amazon              | Searches in Amazon.com                          |
| 2  | Google              | Searches with Google.com                        |
| 3  | Bliquidity          | Information about liquidity in a banking system |
| 4  | Currency Converter  | Does currency conversion                        |
| 5  | Stock Quote         | Latest stock quotes                             |
| 6  | Fast Weather        | Weather forecasts for a city                    |
| 7  | Quote of the Day    | Random daily quotes each day                    |
| 8  | GetJoke             | Provides random jokes                           |
| 9  | Hyperlink Extractor | Extracts all hyperlinks of a web page           |
| 10 | XML Daily Fact      | Daily facts on web services, etc.               |

Cloud-based web services using long-memory time series 13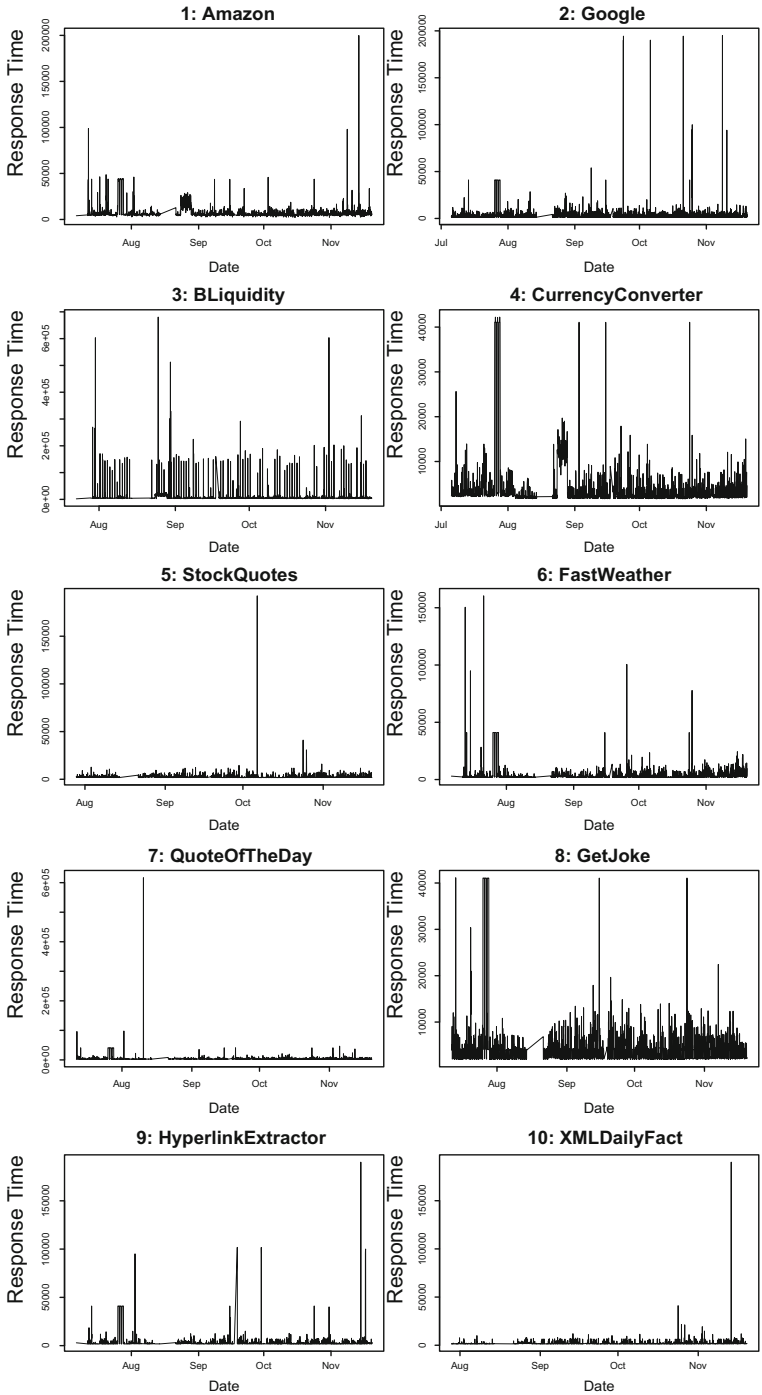

**Fig. 2** Response time of web services (in ms)

Figure 3 shows the sample autocorrelation function (ACF) for all of the 10 services, with lags of up to 96 h (4 days). By examining the ACF diagrams, two types of time series are identifiable:

1. Short-memory or short-range dependent (SRD), in which the ACF decays exponentially quickly to 0 and there is a finite sum. These are services 3, 5, and 10.
2. Long-memory or long-range dependent (LRD), in which the ACF has a power law decay and an infinite sum. This means that the autocorrelation remains non-negligible for a long period time.

Figure 3 also shows that except for services 3, 5 and 10, which are not CPU intensive, the ACF of all of the other services decays slowly. Thus, we can intuitively say that these services have a long memory, and in the next section, we statistically show the existence of this feature by estimating the Hurst exponent. Investigating the causes of long-range dependence in the QoS data requires separate research, but there are other interesting features that are visible in the graphs. The ACF of these services conform to another specific pattern: there are local maxima at lags of 24 (48 for service with a 2-h data collection interval and 12 for service with 30-min data collection interval). This result occurs in LRD time series that have a seasonal component. Please note that because there are different sample rates for services nos. 4 and 7, the same rule works for them.

All of the time series that have long-range dependence belong to services that do information retrieval and/or some processing on the data. Services 1 and 2 search in Amazon and Google, respectively, service 4 retrieves latest currency exchange rates and does the currency conversion, service 5 retrieves weather forecasts for a certain city, service 7 calculates a random number and retrieves a random joke, service 9 retrieves a web page and extracts all hyperlinks of the web page. Response time of each of these web services exhibits LRD.

In contrast, time series that have short-range dependence belong to services that do simple lookups which are not CPU intensive. Service 3 provides information about liquidity in a banking system, service 5 provides latest stock quotes (low frequency data), and service 10 provides daily facts on web services. Response time of each of these web services exhibits SRD.

4.2 Estimating the Hurst exponent

As discussed in the previous sections, the Hurst exponent is used to examine the existence of long-range dependence in a time series. The estimated values of Hurst exponent, according to several estimators, including the aggregated variance, rescaled range, and periodogram, are shown in Table 2.

Considering the shape of the ACF and the seasonality of the data, this variation in estimated value of $H$ (approximately 0.1) is acceptable. Thus, according to the estimated Hurst exponent, except for rows 3, 5 and 10, which are bold values in Table 2, all of the other time series have long-range dependence.

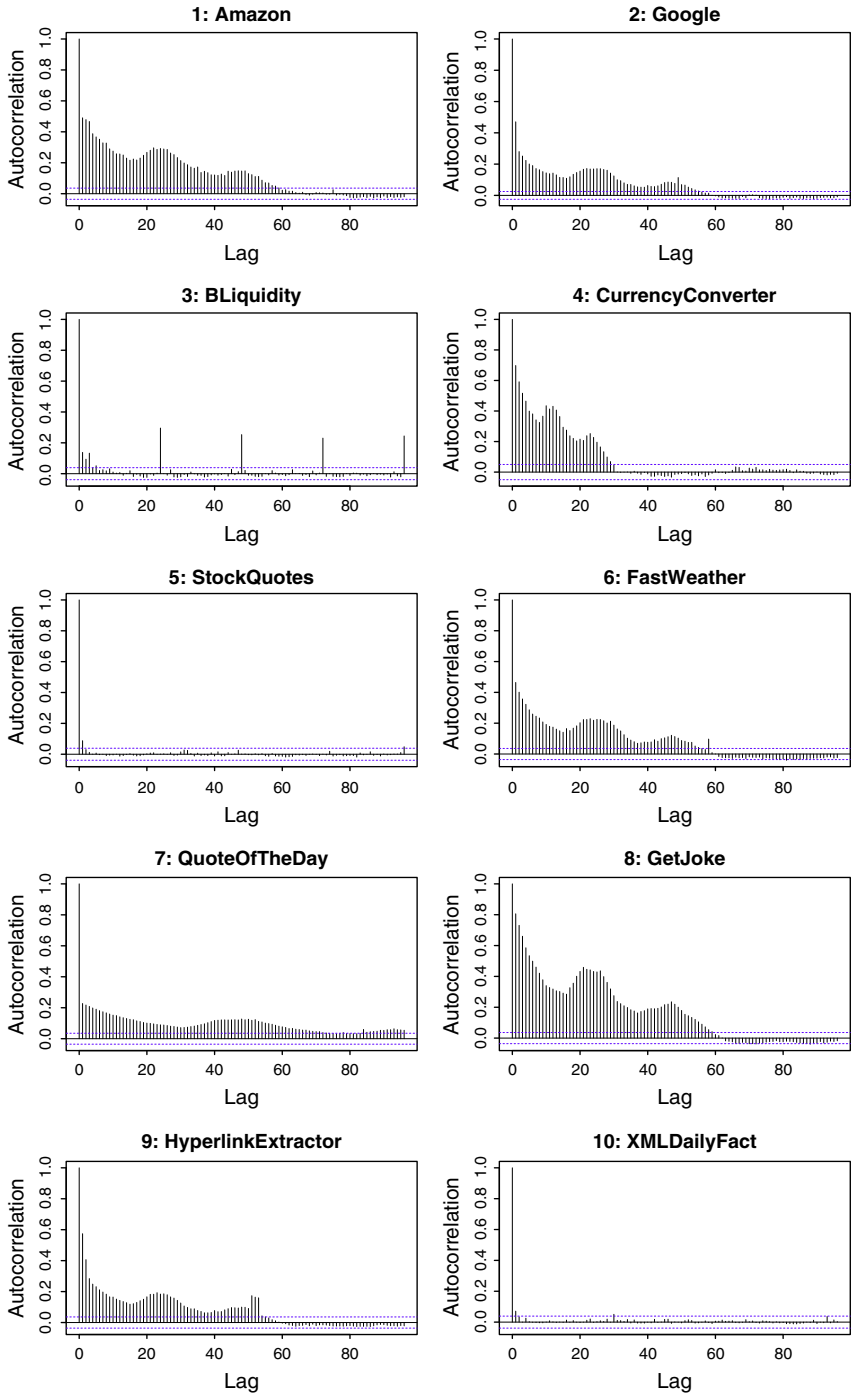

**Fig. 3** Sample autocorrelation for the response time of the 10 web services

Table 2 Estimates of the Hurst exponent

| Hurst exponent service no. | Aggregated variance | R/S | Periodogram |
|---|---|---|---|
| 1 | 0.7601 | 0.7631 | 0.7602 |
| 2 | 0.728 | 0.6629 | 0.7383 |
| 3 | **0.5899** | **0.5759** | **0.668** |
| 4 | 0.8397 | 0.9452 | 0.9768 |
| 5 | **0.5392** | **0.5795** | **0.5087** |
| 6 | 0.8296 | 0.7569 | 0.7968 |
| 7 | 0.7823 | 0.6782 | 0.6069 |
| 8 | 0.8443 | 0.8067 | 1.0605 |
| 9 | 0.7793 | 0.8081 | 0.8893 |
| 10 | **0.5598** | **0.5686** | **0.5053** |

### 4.3 Forecasting results

Forecast error is calculated by subtracting the forecast value from the original measured value: $e_i = y_i - \hat{y}_i$. Forecast accuracy measures are calculated based on $e_i$ and thus they are scale dependent. One of the most commonly used measures is mean absolute error (MAE) that is computed as [10]

$$\text{MAE} = \text{mean}(|e_i|) \qquad (19)$$

Another important measure is mean absolute percentage error (MAPE) that shows mean percentage of error, thus it is scale independent and it is widely used to compare forecast performance of different forecasting methods. MAPE is computed as [10]

$$\text{MAPE} = \text{mean}(|P_i|) \qquad (20)$$

If method 2 provides less error compared to method 1, we calculate improvement percentage according to

$$i = \left(\frac{\text{MAPE}_1 - \text{MAPE}_2}{\text{MAPE}_1}\right) \times 100 \qquad (21)$$

To calculate the precision of the different forecasting mechanisms, four different forecasting methods (naïve, mean, ARIMA and ARFIMA) are used for modeling and forecasting future values.

For calculating forecast performance, we did not confine ourselves to limited experiments with special data. To achieve good accuracy in performance comparison of different methods, we have done cross-validation [29] on all 10 available datasets. We have used a fixed window as the training dataset and calculated the mean of forecast error of each method by comparing forecasts and actual measured values, and we have done this repeatedly by rolling the fixed window from the start to the end of each dataset.

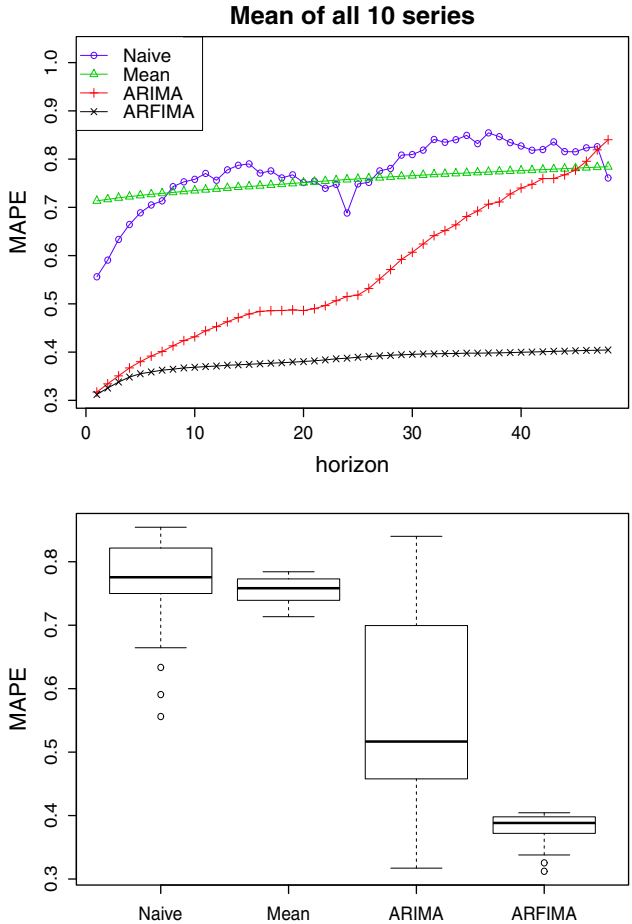

**Fig. 4** Comparing different forecasting methods according to MAPE, calculated over all 10 series

Because all values of response time are strictly positive, we have used logarithmic transformation. This can be viewed as a Box–Cox transformation with $\lambda = 0$, defined as

$$w_t = f(x) = \begin{cases} \log(y_t), & \text{if } \lambda = 0 \\ (y_t^\lambda - 1)/\lambda, & \text{otherwise} \end{cases} \quad (22)$$

We do forecasting on transformed values, and then we transform back these results to the original space. This transformation is done for all of the four forecasting methods.

Figure 4 shows the accuracy of the different forecasting methods according to MAPE for forecast horizons from 1 to 48. Please note that this is calculated over all 10 series. ARFIMA on the mean provides better forecast on all 10 services, and also provides less variant errors that can be seen on the boxplot of the forecast errors. ARFIMA forecast improvement over three other forecasting methods can be seen in Table 3.

**Table 3** ARFIMA forecast improvement over three other forecasting methods according to MAPE (on LRD series)

|  | Naïve (%) | Mean (%) | ARIMA (%) |
|---|---|---|---|
| Mean improvement of ARFIMA compared to … | 44.0 | 44.6 | 37.5 |
| Maximum improvement of ARFIMA compared to … | 48.7 | 54.0 | 57.8 |

Overall, value of forecast error for each method is calculated from more than 29,000 experiments across the 10 different datasets. Results show that for response time of web services that show LRD, ARFIMA provides 37.5 % less error in the mean, and maximum of 57.8 % improvement that occurs in maximum forecast horizon ($h=48$). In our experiments, we have chosen the window size of $4 \times 24 = 96$ and maximum forecast horizon of $2 \times 24 = 48$.

Our experiments are done using a Core i7-M processor. The time needed to fit the ARFIMA model on a window of length 96 and forecast for the next 24 steps, was about 100 ms on this system. We conclude that ARFIMA is fast, and needs small set of measurement history to give acceptable results, thus it is a good choice for forecasting QoS data of web services.

As it can be seen on Fig. 5, ARFIMA shows considerable reduction of forecast error on LRD series. For services 3, 5, and 10 that are not CPU intensive and their measured response time is SRD, difference between forecast error of ARIMA and ARFIMA is limited, but even in these situations, ARFIMA provides slightly better forecasts.

## 5 Conclusions and future works

In this study, we have analyzed measured web service QoS data, and by comparing the results of three different methods, we estimated the Hurst exponent and showed the presence of long-range dependence in measured response time of CPU-intensive web services. Further investigation of ACF for the response times showed the presence of a seasonal part in the data with a period of 24 h. However, ADF test over the data showed that with the large fluctuations of data, the statistical characteristics do not vary much over time, and the data can be considered to be stationary.

Considering the presence of LRD in QoS data, it is possible to use this feature and achieve more exact forecasts for the future values of QoS. Using fractional differencing to model LRD series may not provide superior results in short-term forecasts, but with longer forecast horizons, it can provide better predictions with less error. We have verified the hypothesis by extensive experiments on QoS data. Results of our experiments and comparisons show that in measured response time of web services that exhibit LRD, ARFIMA provides better fit. Also, using standard error measure of MAPE, we have shown that forecasts of ARFIMA have higher accuracy in longer forecast steps compared to the ARIMA, mean and naïve. The other advantage of considering LRD is the achievement of more precise prediction intervals compared with the other methods. Naïve estimation provides an ultra-wide confidence interval that is not usable in practice. In addition, the mean provides a constant confidence

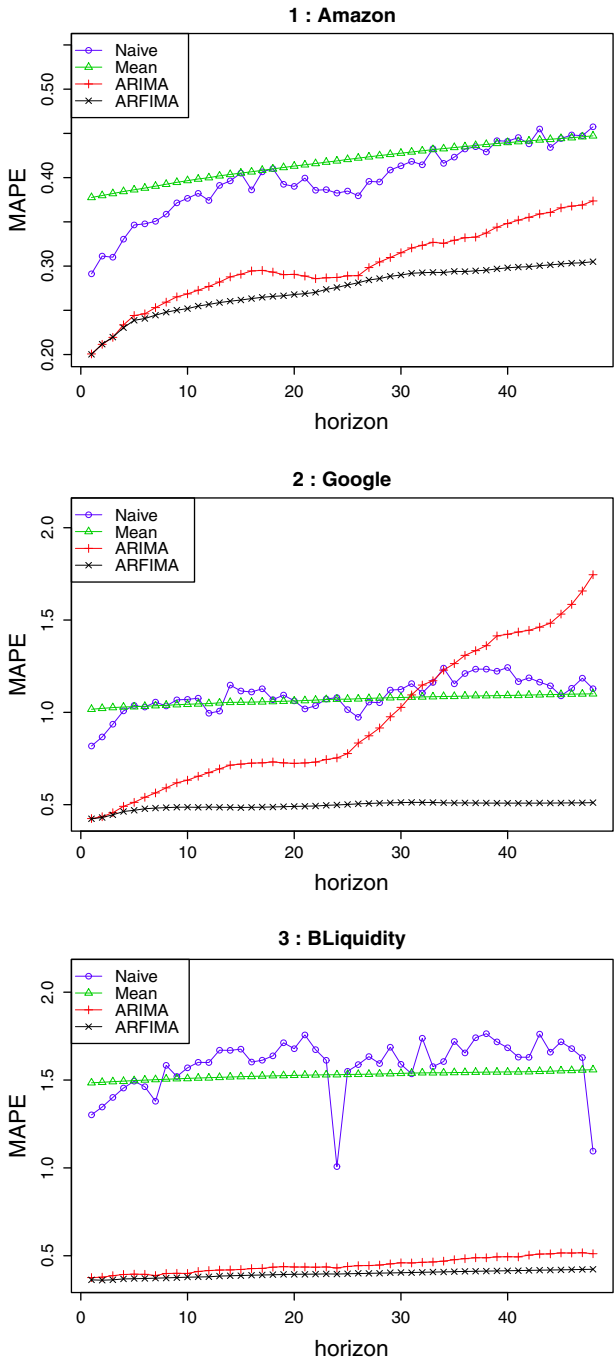

**Fig. 5** Comparing different forecasting methods according to MAPE

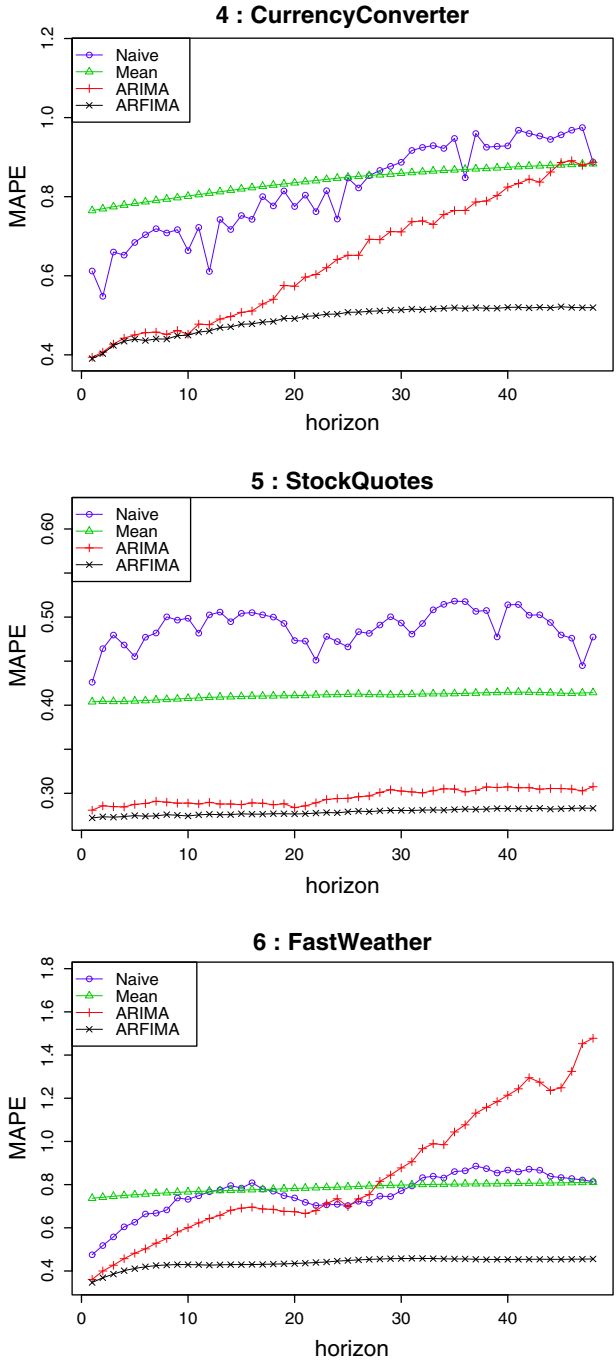

**Fig. 5** continued

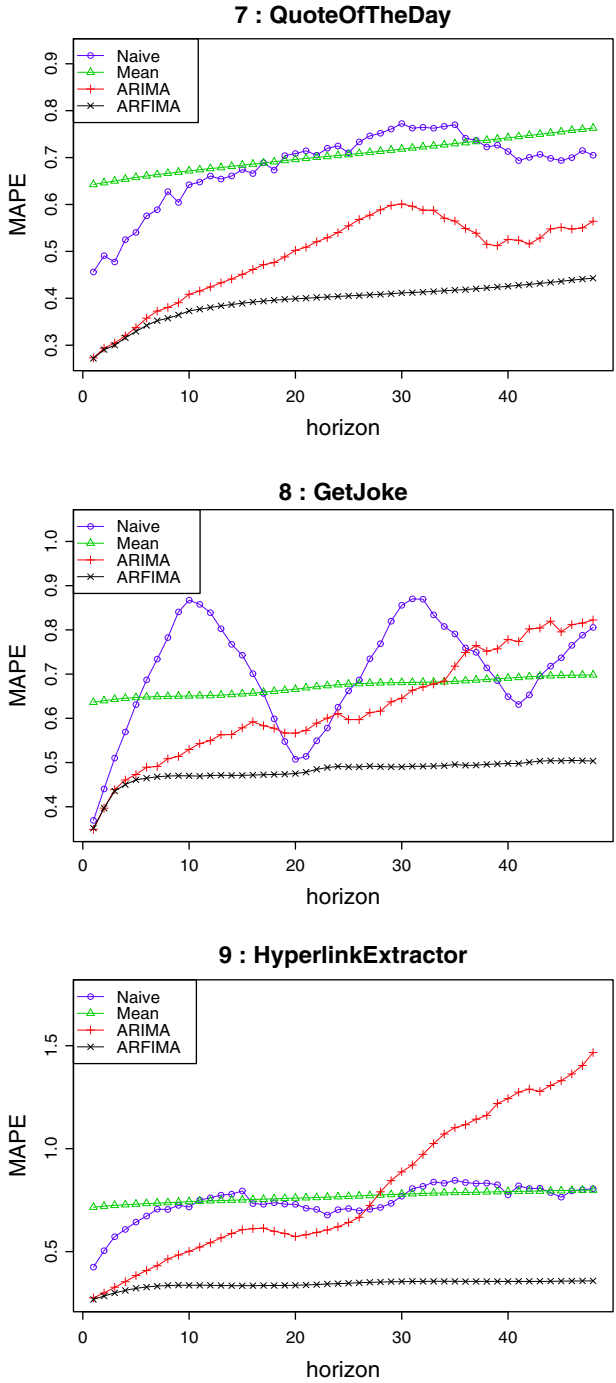

**Fig. 5** continued

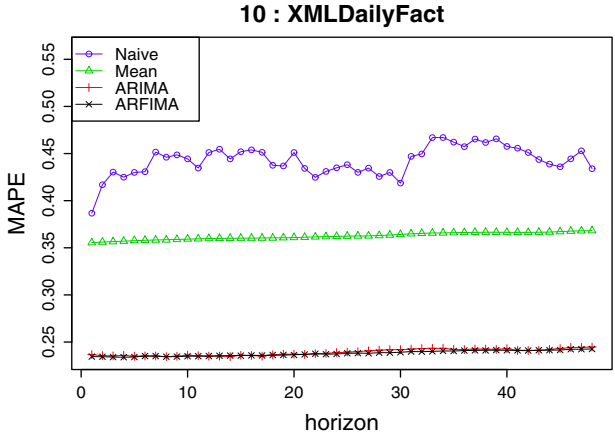

**Fig. 5** continued

interval that is not useful. In the ARIMA method, the confidence interval widens as time passes, and for long-term forecasts it reduces the accuracy. Only ARFIMA can make use of the strong correlation between consecutive values. Thus, using predicted values of this forecast method that makes use of LRD leads to better selection of services from the set of candidates.

In previous related studies, even the latest studies on web services QoS, the memory of these time series has not been investigated, and the existence of long-range dependence was not shown. In this work, by considering LRD feature of QoS data, we have used the ARFIMA forecasting method that makes use of this feature to better forecast future values, and we have compared the accuracy of the ARFIMA to SRD methods.

It is worth noting that further work in this area is needed, and here we provide insight on possible future works. LRD models are not limited to the ARFIMA, thus studying other LRD models like GARMA and FIGARCH, and measuring the forecast performance of these methods on QoS time series can be an area of future research. Also, careful investigations of seasonality in the data, the benefit of using seasonal methods and the effect of lack of data on forecast performance are also needed.

We plan to study more on the roots of LRD in QoS data of web services on the Internet, because the suggested causes of LRD in Internet traffic are not necessarily applicable to the context of web services. In our work, we have observed that response time of services that are not CPU intensive does not have long memory. This observation should be investigated more and confirmed with further experiments. Other than these future works, studying the effect of the provided point forecast and the prediction intervals on the rate of service level agreement (SLA) violation is an important area of the future researches that we aim to follow.